\documentclass[prl, twocolumn]{revtex4}

\usepackage{amsmath,amssymb}  
\usepackage{psfrag}  
\usepackage[dvips]{epsfig}  
\usepackage{epsfig}

\usepackage{theorem}  
\newtheorem{theorem}{Theorem}

\newtheorem{proposition}{Proposition}

\theorembodyfont{\rmfamily}  
  
\theoremstyle{break}  
  
\def\QED{~\rule[-1pt]{5pt}{5pt}\par\medskip}  
  
\newenvironment{proof}{{\bf Proof: \ }}{ \hfill \QED}

\def\su{\text{su}}  
 
\def\SU{\text{SU}}  
 
\begin{document}   
\title{Optimal quantum circuit synthesis from Controlled-$U$
  gates}  
   
\author{Jun Zhang$^{1, 2} $, Jiri Vala$^{2}$,  Shankar Sastry$^1$
and K. Birgitta Whaley$^{2}$} 
  
\affiliation{$^1$Department of Electrical Engineering  
and Computer Sciences, University of California, Berkeley, CA 94720\\  
$^2$Department of Chemistry and Pitzer Center for Theoretical Chemistry, 
University of California, Berkeley, CA 94720} 
 
\date{\today}  
  
\begin{abstract}  
  From a geometric approach, we derive the minimum number of
  applications needed for an arbitrary Controlled-Unitary gate to
  construct a universal quantum circuit. A new analytic construction
  procedure is presented and shown to be either optimal or close to
  optimal.  This result can be extended to improve the efficiency of
  universal quantum circuit construction from any entangling gate.
  Specifically, for both the Controlled-NOT and Double-CNOT gates, we
  develop simple analytic ways to construct universal quantum circuits
  with three applications, which is the least possible.
\end{abstract}   
  
\maketitle  

Construction of a universal quantum circuit, i.e., a circuit that can
implement any arbitrary unitary operation, is of central importance in
the physical applications of quantum computation and quantum
information processing~\cite{Deutsch:85}. Barenco {\it et
  al.}~\cite{Barenco:95} proved the celebrated result that the
Controlled-NOT (CNOT) gate supplemented with single-qubit rotations is
universal, which has become a de facto standard model of quantum
computation. The authors~\cite{Zhang:03} have previously provided a
generality beyond the standard model, namely, an analytic direct route
to simulate any arbitrary two-qubit unitary operation from whatever
entangling gate arises naturally in the physical applications. An
extremely important further issue is the minimum applications required
for a given gate to implement a universal quantum circuit.
In~\cite{Zhang:03}, we provided an upper bound for the
applications of a given entangling gate, i.e., regardless of which
two-qubit gate to be implemented, we can always construct a quantum
circuit with applications of the given gate not exceeding that upper
bound.  However, this upper bound is not tight because it may be
possible to achieve universality with less applications of the given
gate. For example, it was recently shown that just three applications
of the CNOT gate together with local gates are
universal~\cite{Vidal:03, Shende:03b}.

The main contribution of this paper is a more general result for
optimality, namely, the minimum number of applications needed for an
arbitrary Controlled-Unitary (Controlled-$U$) gate to construct a
universal quantum circuit. We focus on the Controlled-$U$ gates
because any entangling two-qubit gate can be used at most twice to
simulate a Controlled-$U$ gate~\cite{Zhang:03}, and these gates are
then widely used as basic building blocks to construct universal
quantum circuits~\cite{DiV:98, Bremner:02}. Our main tool to derive
the minimum upper bound for any Controlled-$U$ gate is the geometric
representation of nonlocal two-qubit gates developed
in~\cite{Zhang:02}, which provides an intuitive approach to this
minimum upper bound. We also obtain a near optimal construction
procedure that implements a universal quantum circuit with either
minimum applications of the given Controlled-$U$ gate, or one
application more than the minimum, depending on the given gate.
Moreover, for the CNOT and Double-CNOT (DCNOT) gate~\cite{Collins:01},
we provide a simple analytic solution to simulate any two-qubit gate
with at most three applications.

\begin{figure}[tb] 
 \begin{center} 
  \psfrag{A1}[][]{$A_1$} 
  \psfrag{A2}[][]{$A_2$} 
  \psfrag{A3}[][]{$A_3$} 
  \psfrag{O}[][]{$O$} 
     \psfrag{C1}[][]{$C_1$} 
   \psfrag{C2}[][]{$C_2$} 
   \psfrag{B1}[][]{$B_1$} 
  \psfrag{C3}[][]{$C_3$} 
 \psfrag{L}[][]{$L$} 
   \psfrag{B3}[][]{$B_3$} 
  \psfrag{B2}[][]{$B_2$} 
  \psfrag{c1}[][]{$c_1$} 
   \psfrag{c2}[][]{$c_2$} 
  \psfrag{c3}[][]{$c_3$} 
 \includegraphics[width=0.7\hsize]{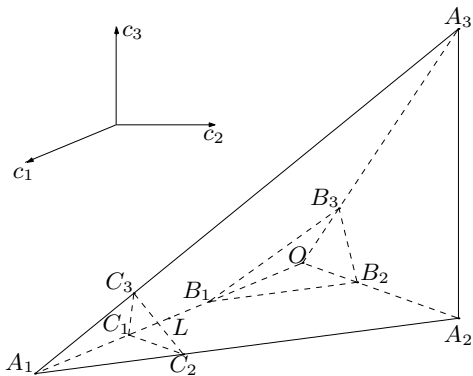}  
 \end{center}
 \caption{Tetrahedron $OA_1A_2A_3$ contains all the local equivalence
   classes of nonlocal gates, where $O([0, 0, 0])$ and $A_1([\pi, 0,
   0])$ both correspond to local gates, $L([\frac\pi{2}, 0, 0])$
   to the CNOT gate, $A_3([\frac\pi{2}, \frac\pi{2}, \frac\pi{2}])$ to
   the SWAP gate, and the Controlled-$U$ gate $U_f=e^{\gamma
     \frac{i}2\sigma_z^1\sigma_z^2}$ to the point $[\gamma, 0, 0]$ on
   $OL$~\cite{Zhang:02}. Tetrahedra $OB_1B_2B_3$ and $A_1C_1C_2C_3$ contain all the
   local equivalence classes of the nonlocal gates that can be
   generated by $n$ applications of $U_f$ with local gates, where
   $B_1=[n\gamma, 0, 0]$, $B_2=[\frac{n\gamma}2, \frac{n\gamma}2, 0]$,
   $B_3=[\frac{n\gamma}3,\frac{n\gamma}3, \frac{n\gamma}3]$,
   $C_1=[\pi-n\gamma, 0, 0]$, $C_2=[\pi-\frac{n\gamma}2,
   \frac{n\gamma}2, 0]$, and $C_3=[\pi-\frac{n\gamma}3,
   \frac{n\gamma}3, \frac{n\gamma}3]$.}
 \label{fig:nr} 
 \end{figure}

\paragraph{Preliminaries}
We first briefly review some relevant background
knowledge~\cite{Makhlin:00,Khaneja:01,Kraus:01,Zhang:02,Zhang:03}.
Two quantum gates $U$, $U_1\in \SU(4)$ are called \emph{locally
  equivalent} if they differ only by local operations: $U= k_1 U_1
k_2$, where $k_1$, $k_2\in \SU(2)\otimes \SU(2)$. Two gates are
locally equivalent if and only if they have identical Makhlin's local
invariants~\cite{Makhlin:00}. From the Cartan decomposition on $\su(4)$,
any two-qubit unitary operation $U\in \SU(4)$ can be written as
\begin{equation}  
  \label{eq:20}  
  U=k_1Ak_2=k_1\cdot e^{ c_1\frac{i}{2}\sigma_x^1\sigma_x^2} \cdot  
e^{c_2\frac{i}{2}\sigma_y^1\sigma_y^2} \cdot  
e^{c_3\frac{i}{2}\sigma_z^1\sigma_z^2}\cdot k_2,  
\end{equation} 
where $\sigma_{\alpha}^1\sigma_{\alpha}^2 = \sigma_{\alpha}\otimes
\sigma_{\alpha}$, $\sigma_\alpha$ are the Pauli matrices, and $k_1$,
$k_2\in {\SU(2)\otimes \SU(2)}$ are local gates. 
In~\cite{Zhang:02} we found that the local equivalence classes of 
two-qubit gates are in one-to-one correspondence with the points in 
the tetrahedron $OA_1A_2A_3$ shown in Fig.~\ref{fig:nr}, except on its base. 
For a general two-qubit gate $U$ in Eq.~\eqref{eq:20}, this geometric
representation defines a set of parameters $c_j$ satisfying
$\pi-c_2\ge c_1\ge c_2\ge c_3\ge 0$. 

Consider an arbitrary single-qubit gate $U=\exp(n_x i\sigma_x+n_y
i\sigma_y+n_z i\sigma_z)$. The Controlled-$U$ operation $U_f$ derived from
this gate can be written as:
\begin{eqnarray}
  \label{eq:41}
U_f=(I\otimes e^{-\gamma\frac{i}{2}\sigma_z}U_1^\dag)\cdot
e^{\gamma\frac{i}2\sigma_z^1\sigma_z^2}\cdot(I\otimes U_1),
\end{eqnarray}
where $\gamma=\sqrt{n_x^2+n_y^2+n_z^2}$, and $U_1$ is a single-qubit
gate given by Proposition 3 of~\cite{Zhang:03}. By definition,
$e^{\gamma\frac{i}2\sigma_z^1\sigma_z^2}$ is locally equivalent to
a Controlled-$U$ gate. Therefore, without loss of generality, we can use
$U_f=e^{\gamma\frac{i}2\sigma_z^1\sigma_z^2}$ to denote any
Controlled-$U$ gate. Since
$e^{(\pi-\gamma)\frac{i}2\sigma_z^1\sigma_z^2}$ is locally equivalent
to $e^{\gamma\frac{i}2\sigma_z^1\sigma_z^2}$, we can always take the
parameter $\gamma\in (0, \frac\pi{2}]$.  Specifically, when
$\gamma=\frac\pi{2}$, $U_f$ is equal to the CNOT gate.
  
\paragraph{Minimum upper bound for any Controlled-$U$ gate}
We have previously provided an upper bound for a given entangling gate
to implement a universal quantum circuit~\cite{Zhang:03}.  For a
Controlled-$U$ gate $U_f=e^{\gamma\frac{i}2\sigma_z^1\sigma_z^2}$,
this upper bound is $6\lceil \frac{\pi}{4\gamma}\rceil$, where the
ceiling function $\lceil x \rceil$ is defined as a function that
rounds $x$ to the nearest integer towards infinity. This upper bound
is not a tight one.  We now use a geometric approach to show that the
minimum upper bound for a Controlled-$U$ gate is $\lceil
\frac{3\pi}{2\gamma}\rceil$.

We know that any Controlled-$U$ gate $U_f$ corresponds to a point on
the line segment $OL$ as shown in Fig.~\ref{fig:nr}. We now study the
set of all the nonlocal gates that can be implemented by $n$
applications of $U_f$. We will first analyze the case $n\ge 3$, and
then the case $n=2$. The following theorem shows that all
gates that can be simulated by $n$ ($\ge 3$) applications of $U_f$
together with local gates constitute two congruent tetrahedra in the
tetrahedron $OA_1A_2A_3$, which is the geometric representation of all
the nonlocal two-qubit operations.

\begin{theorem}
\label{thm:nr}
For a Controlled-$U$ gate
$U_f=e^{\gamma\frac{i}2\sigma_z^1\sigma_z^2}$, every gate generated by
$n$ ($\ge 3$) applications of $U_f$ together with local gates is
locally equivalent to a gate $e^{ c_1\frac{i}{2}\sigma_x^1\sigma_x^2}
e^{ c_2\frac{i}{2}\sigma_y^1\sigma_y^2}
e^{c_3\frac{i}{2}\sigma_z^1\sigma_z^2}$, with the parameters $c_j$
satisfying either $0\le c_1+c_2+c_3\le n\gamma$ or $c_1-c_2-c_3\ge
\pi-n\gamma$.
\end{theorem}
 
The proof of this theorem is mathematically complex and will be
presented elsewhere. Theorem~\ref{thm:nr} tells us
that all the gates that can be generated by $n$ applications of $U_f$
with local gates can be represented by two tetrahedra $OB_1B_2B_3$ and
$A_1C_1C_2C_3$ in Fig.~\ref{fig:nr}. Note that these two tetrahedra
are congruent, and the equations describing the faces $B_1B_2B_3$ and
$C_1C_2C_3$ are $c_1+c_2+c_3=n\gamma$ and $c_1-c_2-c_3= \pi-n\gamma$,
respectively.  These two faces are the boundaries of all those points
that can be generated by $n$ applications of $U_f$.

It is clear that as $n$ grows, each of these two tetrahedra
$OB_1B_2B_3$ and $A_1C_1C_2C_3$ expands with consecutive faces of each
tetrahedron remaining parallel. To obtain the minimum number of
applications needed for a given Controlled-$U$ gate $U_f$ to implement
any arbitrary two-qubit operation, we only need to find the least
integer $n$ such that the union of the two tetrahedra $OB_1B_2B_3$ and
$A_1C_1C_2C_3$ can cover the whole tetrahedron $OA_1A_2A_3$ as $n$
grows. Since this is convex, we can further restrict our attention to
covering all its vertices. As seen from Fig.~\ref{fig:nr}, this is
equivalent to the condition that one of the two tetrahedra contains
the point $A_3([\frac\pi{2}, \frac\pi{2}, \frac\pi{2}])$, i.e., the
SWAP gate.  From Theorem~\ref{thm:nr}, we only require that
$n\gamma\ge \frac{3\pi}2$, which leads to $n=\lceil
\frac{3\pi}{2\gamma} \rceil$.  This provides the minimum upper bound
for an arbitrary Controlled-$U$ gate to implement a universal quantum
circuit, and is summarized in the following theorem.

\begin{theorem}
\label{prop:minimum}
For an arbitrary Controlled-$U$ gate
$U_f=e^{\gamma\frac{i}2\sigma_z^1\sigma_z^2}$, the minimum
applications required to implement any arbitrary two-qubit gate
together with local gates is $\lceil\frac{3\pi}{2\gamma}\rceil$.
\end{theorem}

\begin{figure}[tb] 
\begin{center} 
 \psfrag{MN}[][]{$\frac\pi{2}$} 
 \psfrag{NM}[][]{$\frac{\pi}{2}$} 
 \psfrag{MM}[][]{$\frac{3\pi}{8}$} 
 \psfrag{NN}[][]{$\frac{3\pi}{10}$} 
 \psfrag{M}[][]{$\frac{\pi}{4}$}
 \psfrag{N}[][]{$\frac{3\pi}{14}$}
 \psfrag{N1}[][]{$\frac{3\pi}{16}$}
\psfrag{2}[][]{$2$} 
\psfrag{4}[][]{$4$} 
\psfrag{6}[][]{$6$} 
\psfrag{8}[][]{$8$} 
\psfrag{10}[][]{$10$} 
\psfrag{12}[][]{$12$} 
\psfrag{14}[][]{$14$} 
\psfrag{r}[][]{$\gamma$} 
\psfrag{0}[][]{$0$} 
\includegraphics[width=0.8\hsize]{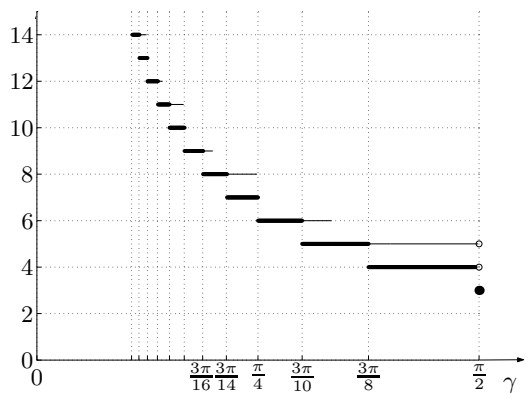}  
\end{center} 
\caption{Upper bound of applications needed for an arbitrary
  Controlled-$U$ gate $U_f=e^{\gamma\frac{i}2\sigma_z^1\sigma_z^2}$
  to construct a universal quantum circuit. Thick lines:
  minimum; thin lines: our constructive approach.}
 \label{fig:upperbound} 
\end{figure}

In Fig.~\ref{fig:upperbound}, the minimum upper bound for any
Controlled-$U$ gate $U_f=e^{\gamma\frac{i}2\sigma_z^1\sigma_z^2}$ is
shown as a function of $\gamma$, and depicted by thick lines. The thin
lines represent the number of applications needed by a near optimal
construction procedure we present below. Note that the single point at
$\gamma=\frac\pi{2}$ with value 3 indicates that three applications of
the CNOT gate with local gates suffice to implement any arbitrary
two-qubit gate. The CNOT gate is therefore the most efficient gate
among all the Controlled-$U$ gates.

\paragraph{Near optimal universal quantum circuit}
In real physical applications, it is desirable to have a constructive
procedure to implement a universal quantum circuit.  At this time,
there is no explicit way to construct a universal quantum circuit that
exactly achieves the minimum upper bound for an arbitrary
Controlled-$U$ gate $U_f$.  However, we have found a construction
procedure for a near optimal universal quantum circuit from an
arbitrary Controlled-$U$ gate $U_f=e^{\gamma\frac{i}2
  \sigma_z^1\sigma_z^2}$ combined with local gates.  Depending on the
value of $\gamma$, the upper bound of this construction is either
equal to the minimum or just one more than the minimum applications of
$U_f$ as shown in Fig.~\ref{fig:upperbound}.

An arbitrary two-qubit operation $U\in \SU(4)$ can be written as in
Eq.~\eqref{eq:20}, with the parameters $c_j$ in the tetrahedron
$OA_1A_2A_3$.  Since we have easy access to all the local
gates~\cite{Bremner:02, Zhang:03}, we only need to implement the
nonlocal part $A$ in Eq.~\eqref{eq:20}. We do this in the following
two steps:

\noindent
{\bf 1.} Apply $e^{\gamma\frac{i}2 \sigma_z^1\sigma_z^2}$ at most
  $\lceil \frac{\pi}{2\gamma}\rceil$ times to simulate the third
  component $e^{c_3\frac{i}2 \sigma_z^1\sigma_z^2}$ of $A$
  (See Proposition 2, \cite{Zhang:03}).

\noindent
{\bf 2.} Apply $e^{\gamma\frac{i}2 \sigma_z^1\sigma_z^2}$ at most
  $\lceil \frac\pi{\gamma}\rceil$ times to simulate the first two
  components $e^{c_1\frac{i}2 \sigma_x^1\sigma_x^2} \cdot
  e^{c_2\frac{i}2 \sigma_y^1\sigma_y^2}$ of $A$.
  (Theorem~\ref{prop:2cu}).
  
  The first step follows directly from Proposition 2
  in~\cite{Zhang:03}. The construction procedure therein takes at most
  $\lceil\frac\pi{2\gamma}\rceil$ applications when $\gamma\in(0,
  \frac\pi{2})$, and only two applications when $\gamma=\frac\pi{2}$,
  i.e., for the CNOT gate. We therefore only need to realize the
  second step. The next theorem identifies all nonlocal gates that can
  be implemented by two Controlled-$U$ gates together with local
  gates.

\begin{theorem}  
  \label{prop:2cu}  
  Given two Controlled-$U$ gates $e^{\gamma_1
    \frac{i}2\sigma_z^1\sigma_z^2}$ and
  $e^{\gamma_2\frac{i}2\sigma_z^1\sigma_z^2}$ with $\gamma_1$,
  $\gamma_2\in (0, \frac\pi{2}]$, all the local equivalence classes
  of two-qubit gates that can be implemented by these two gates
  together with local gates can be described as
  $e^{c_1\frac{i}2\sigma_x^1\sigma_x^2}\cdot
  e^{c_2\frac{i}2\sigma_y^1\sigma_y^2}$ with $0\le c_1+c_2\le
  \gamma_1+\gamma_2$. Furthermore, we can implement such a gate
by the following quantum circuit:
\setlength{\unitlength}{0.16cm}  
\begin{center}  
\scriptsize
\begin{picture}(37,7)
\put(0,0){\begin{picture}(37,7)
\put(0,1.5){\line(1,0){3}}  
\put(0,5.5){\line(1,0){3}}  
\put(3,0){\thicklines\framebox(9,7)[c]{$e^{\gamma_1\frac{i}2\sigma_z^1\sigma_z^2}$}}  
\put(12,1.5){\line(1,0){3}}  
\put(12,5.5){\line(1,0){3}}
\put(15,0){\framebox(7,3)[c]{$e^{\beta_2 \frac{i}2 \sigma_y}$}}
\put(15,4){\framebox(7,3)[c]{$e^{\beta_1 \frac{i}2 \sigma_y}$}}
\put(22,1.5){\line(1,0){3}}  
\put(22,5.5){\line(1,0){3}}
\put(25,0){\thicklines\framebox(9,7)[c]{$e^{\gamma_2\frac{i}2\sigma_z^1\sigma_z^2}$}}  
\put(34,1.5){\line(1,0){3}}  
\put(34,5.5){\line(1,0){3}}
\end{picture}}
\end{picture}  
\end{center}  
where $\cos\beta_1$ and $\cos\beta_2$ are the two roots of the quadratic equation:
\begin{eqnarray}
\label{eq:88}
\aligned
\sin&\gamma_1\sin\gamma_2\ x^2 +
\big(\cos^2c_1+\cos^2c_2-\cos^2\gamma_1-\cos^2\gamma_2\\
&+2(\cos\gamma_1\cos\gamma_2-\cos c_1\cos c_2)
\cos(\gamma_1-\gamma_2)\big)^{\frac12}x\\
&+\cos\gamma_1\cos\gamma_2-\cos c_1\cos c_2=0.
\endaligned
\end{eqnarray}
\end{theorem} 

 \begin{figure}[tb] 
 \begin{center} 
  \psfrag{XA}[][]{$\frac\pi{2}$} 
  \psfrag{XB}[][]{$\gamma_1+\gamma_2$} 
  \psfrag{XC}[][]{$\pi$} 
  \psfrag{YA}[][]{$\frac\pi{2}$} 
  \psfrag{YB}[][]{$\pi$} 
  \psfrag{A2}[][]{$A_2$} 
  \psfrag{A1}[][]{$A_1$} 
  \psfrag{0}[][]{$O$} 
    \psfrag{c1}[][]{$c_1$} 
   \psfrag{c2}[][]{$c_2$} 
 \includegraphics[width=0.7\hsize]{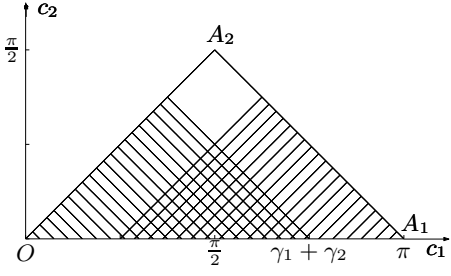}  
 \end{center}
 \caption{Nonlocal gates that can be generated by two given Controlled-$U$ gates 
   $e^{\gamma_1 \frac{i}2\sigma_z^1\sigma_z^2}$ and
   $e^{\gamma_2\frac{i}2\sigma_z^1\sigma_z^2}$.}
 \label{fig:base} 
 \end{figure} 
 
 See Appendix for a proof. This theorem can be illustrated by
 Fig.~\ref{fig:base}, in which the triangle $OA_1A_2$ is the base of
 the tetrahedron $OA_1A_2A_3$, and the Controlled-$U$ gates
 $e^{\gamma_1 \frac{i}2\sigma_z^1\sigma_z^2}$ and $e^{\gamma_2
   \frac{i}2\sigma_z^1\sigma_z^2}$ correspond to points $[\gamma_1,
 0]$ and $[\gamma_2, 0]$ on $OA_1$, respectively.  The nonlocal gates
 that can be generated by these two Controlled-$U$ gates are shown as
 the shaded area in Fig.~\ref{fig:base}. Since the gate $[c_1, c_2,
 0]$ is locally equivalent to the gate $[\pi-c_1, c_2, 0]$, the shaded
 area consists of two symmetric triangles. (Note that Proposition 2
 in~\cite{Zhang:03} is a special case of this theorem by setting
 $\beta_1=2\pi$ and $\gamma_1=\gamma_2$.) When
 $\gamma_1=\gamma_2=\frac\pi{2}$, i.e., both gates are CNOT gates, the
 above quantum circuit can implement any gate in the triangle
 $OA_1A_2$. In other words, two applications of the CNOT gate can
 implement those two-qubit gates that are located on the base of the
 tetrahedron $OA_1A_2A_3$ and only those gates. This result was also
 implied in~\cite{Vidal:03}.
 
 Since the second step of the procedure is indeed equivalent to
 implementing any gate in the triangle $OA_1A_2$, we can now realize
 it by using Theorem~\ref{prop:2cu}. From a given Controlled-$U$ gate
 $U_f=e^{\gamma \frac{i}2\sigma_z^1\sigma_z^2}$, it is easy to obtain
 an $n$-fold product gate $e^{n\gamma \frac{i}2\sigma_z^1\sigma_z^2}$
 by $n$ applications of $U_f$. We then take $\gamma_1=n\gamma$ and
 $\gamma_2=m\gamma$.  From Theorem~\ref{prop:2cu}, to ensure that
 $e^{n\gamma \frac{i}2\sigma_z^1\sigma_z^2}$ and $e^{m\gamma
   \frac{i}2\sigma_z^1\sigma_z^2}$ can simulate any gate in the
 triangle $OA_1A_2$, we only require that the shaded area in
 Fig.~\ref{fig:base} covers the point $A_2$. This is equivalent to
 $(m+n)\gamma \ge\pi$, whence $m+n=\lceil \frac\pi{\gamma}\rceil$. We
 can therefore choose any positive integers $m$ and $n$, as long as
 they satisfy this equality. Moreover, the parameters $\beta_1$ and
 $\beta_2$ of the local gates can be determined by solving
 Eq.~\eqref{eq:88}. Hence we can explicitly simulate any nonlocal gate
 $e^{c_1\frac{i}2 \sigma_x^1\sigma_x^2} e^{c_2\frac{i}2
   \sigma_y^1\sigma_y^2}$ by applying the Controlled-$U$ gate
 $U_f=e^{\gamma \frac{i}2\sigma_z^1\sigma_z^2}$ at most $\lceil
 \frac\pi{\gamma}\rceil$ times.
 
 Combining these two steps together, for a given Controlled-$U$ gate
 $U_f=e^{\gamma \frac{i}2\sigma_z^1\sigma_z^2}$, the constructive
 approach presented above needs at most
 $\lceil\frac\pi{\gamma}\rceil+\lceil\frac\pi{2\gamma}\rceil$
 applications for the case $\gamma\in(0, \frac\pi{2})$, or $4$
 applications for the case $\gamma=\frac\pi{2}$, to implement any
 arbitrary two-qubit operation.  In Fig.~\ref{fig:upperbound}, the
 upper bound of this construction procedure is shown as thin lines. It
 is evident that our procedure is near optimal -- it implements a
 universal quantum circuit with either minimum possible applications
 of $U_f$, or one more than the minimum.
 
 In~\cite{Zhang:03} we provided an upper bound of
 $6\lceil\frac\pi{4\gamma}\rceil$ applications for an arbitrary
 Controlled-$U$ gate $U_f$.  Since
 $\lceil\frac\pi{\gamma}\rceil+\lceil\frac\pi{2\gamma}\rceil\le
 6\lceil\frac\pi{4\gamma}\rceil$, it is clear that the construction
 procedure presented here is more optimal. Furthermore, since $U_f$ is
 a basic building block for implementing a universal quantum circuit
 from any arbitrary entangling gate, we also obtain a much more
 efficient route to this more general goal~\cite{Barenco:95, DiV:98,
  Bremner:02, Zhang:03}.

\paragraph{Universal quantum circuit from three CNOT or DCNOT gates}
The explicit construction procedure presented above requires four
applications of the CNOT gate to implement any arbitrary two-qubit
gate. From Theorem~\ref{prop:minimum}, we know that the minimum upper
bound for the CNOT gate is three (see also Fig.~\ref{fig:upperbound}).
Since the CNOT gate with local gates are widely adopted as the
standard model of universal quantum computation, it is especially
important to find an attractive construction with a minimum number of
applications. Recent work has provided constructions with three
applications of CNOT~\cite{Vidal:03, Shende:03b}. We have found the
following simple analytic route to construct a universal quantum
circuit from three applications of the CNOT gate with local gates.

\begin{proposition}
\label{prop:cnot}
The following quantum circuit is locally equivalent to a generic
nonlocal gate $A=e^{ c_1\frac{i}{2}\sigma_x^1\sigma_x^2}
e^{c_2\frac{i}{2}\sigma_y^1\sigma_y^2}
e^{c_3\frac{i}{2}\sigma_z^1\sigma_z^2}$:
\begin{widetext}
\setlength{\unitlength}{0.16cm}
\scriptsize
\begin{center}  
\begin{picture}(70,7)  
\put(0,1.5){\line(1,0){3}}  
\put(0,5.5){\line(1,0){3}}  
\put(3,0){\thicklines\framebox(9,7)[c]{$e^{\frac\pi{2}\frac{i}2\sigma_z^1\sigma_z^2}$}}  
\put(12,1.5){\line(1,0){3}}  
\put(12,5.5){\line(1,0){3}}
\put(15,0){\framebox(7,3)[c]{$e^{\frac\pi{2} \frac{i}2 \sigma_y}$}}
\put(15,4){\framebox(7,3)[c]{$e^{c_1 \frac{i}2 \sigma_y}$}}
\put(22,1.5){\line(1,0){3}}  
\put(22,5.5){\line(1,0){3}}
\put(25,0){\thicklines\framebox(9,7)[c]{$e^{\frac\pi{2}\frac{i}2\sigma_z^1\sigma_z^2}$}}  
\put(34,1.5){\line(1,0){3}}  
\put(34,5.5){\line(1,0){8.5}}
\put(37,0){\framebox(18,3)[c]{$e^{\frac\pi{2} \frac{i}2 (\sin
      c_3\sigma_x+\cos c_3\sigma_y)}$}}
\put(42.5,4){\framebox(7,3)[c]{$e^{c_2 \frac{i}2 \sigma_x}$}}
\put(55,1.5){\line(1,0){3}}  
\put(49.5,5.5){\line(1,0){8.5}}
\put(58,0){\thicklines\framebox(9,7)[c]{$e^{\frac\pi{2}\frac{i}2\sigma_z^1\sigma_z^2}$}}
\put(67,1.5){\line(1,0){3}}  
\put(67,5.5){\line(1,0){3}}
\end{picture}  
\end{center}  
\end{widetext}
\end{proposition}
\begin{proof}
  By direct algebraic computation, we can show that Makhlin's local
  invariants~\cite{Makhlin:00} of the above quantum circuit are
  identical to those of the nonlocal gate $A$ (See Eq. (25)
  in~\cite{Zhang:02}). Therefore this quantum circuit implements the
  nonlocal gate $A$.
\end{proof}

Moreover, we have a similar result for the DCNOT gate,
which is defined as the quantum gate performing the operation:
$|m\rangle \otimes |n\rangle \to |n\rangle \otimes |m\oplus
n\rangle$~\cite{Collins:01}. It is easy to prove that the DCNOT gate
is locally equivalent to the gate
$e^{\frac\pi{2}\frac{i}2\sigma_x^1\sigma_x^2+
  \frac\pi{2}\frac{i}2\sigma_y^1\sigma_y^2}$, which corresponds to
$A_2$($[\frac\pi{2}, \frac\pi{2}, 0]$) in Fig.~\ref{fig:nr}. Note
that this is not a Controlled-$U$ gate. 

\begin{proposition}
The following quantum circuit is locally equivalent to a generic
nonlocal gate $A=e^{ c_1\frac{i}{2}\sigma_x^1\sigma_x^2}
e^{c_2\frac{i}{2}\sigma_y^1\sigma_y^2}
e^{c_3\frac{i}{2}\sigma_z^1\sigma_z^2}$:
\begin{widetext}
\setlength{\unitlength}{0.16cm} 
\scriptsize 
\begin{center}  
\begin{picture}(92,7)  
\put(0,1.5){\line(1,0){3}}  
\put(0,5.5){\line(1,0){3}}  
\put(3,0){\thicklines\framebox(16,7)[c]{$e^{\frac\pi{2}\frac{i}2\sigma_x^1\sigma_x^2+
\frac\pi{2}\frac{i}2\sigma_y^1\sigma_y^2}$}}  
\put(19,1.5){\line(1,0){3}}  
\put(19,5.5){\line(1,0){3}}
\put(22,0){\framebox(10,3)[c]{$e^{\frac\pi{2} \frac{i}2 \sigma_y}$}}
\put(22,4){\framebox(10,3)[c]{$e^{(\frac\pi{2}-c_1) \frac{i}2 \sigma_y}$}}
\put(32,1.5){\line(1,0){3}}  
\put(32,5.5){\line(1,0){3}}
\put(35,0){\thicklines\framebox(16,7)[c]{$e^{\frac\pi{2}\frac{i}2\sigma_x^1\sigma_x^2+
\frac\pi{2}\frac{i}2\sigma_y^1\sigma_y^2}$}}  
\put(51,1.5){\line(1,0){3}}  
\put(51,5.5){\line(1,0){3}}
\put(54,4){\framebox(16,3)[c]{$ e^{\frac\pi{2}
    \frac{i}2 \sigma_y} e^{(\frac{\pi}{2}-c_3)\frac{i}2\sigma_z} $}}
\put(54,0){\framebox(16,3)[c]{$ e^{(\frac{3\pi}{2}-c_2)\frac{i}2\sigma_y}
e^{\frac\pi{2}\frac{i}2 \sigma_z} $}}
\put(70,1.5){\line(1,0){3}}  
\put(70,5.5){\line(1,0){3}}
\put(73,0){\thicklines\framebox(16,7)[c]{$e^{\frac\pi{2}\frac{i}2\sigma_x^1\sigma_x^2+
\frac\pi{2}\frac{i}2\sigma_y^1\sigma_y^2}$}}  
\put(89,1.5){\line(1,0){3}}  
\put(89,5.5){\line(1,0){3}}
\end{picture}  
\end{center}
\end{widetext}  
\end{proposition}

This proposition can also be proved by direct algebraic computation of
Makhlin's invariants, as for Proposition~\ref{prop:cnot}.

\paragraph{Conclusion}
In summary, we have found the minimum upper bound to construct
a universal quantum circuit from any Controlled-$U$ gate together with
local gates. This minimum upper bound depends only on the single
Controlled-$U$ parameter $\gamma$, as shown in
Fig.~\ref{fig:upperbound}. It shows that among all the
Controlled-$U$ gates, the CNOT gate is the most efficient, a fact not
evident from the previous upper bound result in~\cite{Zhang:03}.
A new explicit construction of universal quantum circuits from a given
Controlled-$U$ gate was provided and shown to be close to optimal,
i.e., it implements a universal quantum circuit with either minimum
applications, or one more than the minimum. For the CNOT gate, this
gives four applications, which is one more than optimal and than other
recent results~\cite{Vidal:03, Shende:03b}. We then developed simple
analytic ways for both the CNOT and DCNOT (not a Controlled-$U$
) gate to construct universal quantum circuits with exactly three
applications, which is the least possible.

\begin{acknowledgments}
We thank the NSF for financial support under ITR Grant No.  
EIA-0205641.
\end{acknowledgments}

\bibliographystyle{apsrev} 

\begin{widetext}
\appendix*
\section{Appendix. Proof of Theorem~\ref{prop:2cu}}
\begin{proof}
  A general two-qubit quantum circuit that consists of two
  Controlled-$U$ gates $e^{\gamma_1 \frac{i}2\sigma_z^1\sigma_z^2}$
  and $e^{\gamma_2\frac{i}2\sigma_z^1\sigma_z^2}$ together with local
  gates can be described as
\begin{equation}
\label{eq:1}
  e^{\gamma_2\frac{i}2\sigma_z^1\sigma_z^2}\cdot
(k_1\otimes k_2)\cdot
e^{\gamma_1 \frac{i}2\sigma_z^1\sigma_z^2}.
\end{equation}
Recall that the local gates $k_1$ and $k_2$ can be written in Euler's
ZYZ decomposition as:
\begin{eqnarray}
\label{eq:2}
\aligned
  k_1&=e^{\alpha_1 i \sigma_z}\cdot e^{\beta_1 i \sigma_y}\cdot
e^{\gamma_1 i \sigma_z},\\
  k_2&=e^{\alpha_2 i \sigma_z}\cdot e^{\beta_2 i \sigma_y}\cdot
e^{\gamma_2 i\sigma_z}.
\endaligned
\end{eqnarray}
Substituting Eq.~\eqref{eq:2} into Eq.~\eqref{eq:1}, and taking into
account the fact that $\sigma_z^1$ and $\sigma_z^2$ both commute with
$\sigma_z^1\sigma_z^2$, we obtain the following quantum circuit that
is locally equivalent to Eq.~\eqref{eq:1}:
\vspace{0.2cm}
\setlength{\unitlength}{0.16cm}  
\begin{center}  
\scriptsize
\begin{picture}(37,7)  
\put(0,1.5){\line(1,0){3}}  
\put(0,5.5){\line(1,0){3}}  
\put(3,0){\thicklines\framebox(9,7)[c]{$e^{\gamma_1\frac{i}2\sigma_z^1\sigma_z^2}$}}  
\put(12,1.5){\line(1,0){3}}  
\put(12,5.5){\line(1,0){3}}
\put(15,0){\framebox(7,3)[c]{$e^{\beta_2 \frac{i}2 \sigma_y}$}}
\put(15,4){\framebox(7,3)[c]{$e^{\beta_1 \frac{i}2 \sigma_y}$}}
\put(22,1.5){\line(1,0){3}}  
\put(22,5.5){\line(1,0){3}}
\put(25,0){\thicklines\framebox(9,7)[c]{$e^{\gamma_2\frac{i}2\sigma_z^1\sigma_z^2}$}}  
\put(34,1.5){\line(1,0){3}}  
\put(34,5.5){\line(1,0){3}}
\end{picture}  
\end{center}  
\vspace{0.2cm} We want to find all the nonlocal gates that can be
generated by the above quantum circuit by tuning the parameters
$\beta_1$ and $\beta_2$ of the local gates. Following the procedure
in~\cite{Makhlin:00}, we find that Makhlin's local
invariants for this quantum circuit are
\begin{eqnarray}
  \label{eq:4}
\aligned
g_1&=\cos r_1 \cos r_2-\sin r_1 \sin r_2 \cos\beta_1 \cos\beta_2,\\
g_2&=0,\\
g_3&=2(\cos\beta_1+\cos\beta_2)^2\sin^2\gamma_1\sin^2\gamma_2
+2\cos^2\gamma_1+2\cos^2\gamma_2-1\\
&\quad
-4\cos\beta_1\cos\beta_2\sin\gamma_1\sin\gamma_2\cos(\gamma_1-\gamma_2).\\ 
\endaligned  
\end{eqnarray}
From~\cite{Zhang:02}, we know that these Makhlin's invariants can also
be written as functions of the parameters $c_j$ in the geometric representation:
\begin{eqnarray}
  \label{eq:24}
  \aligned
g_1&=\cos c_1\cos c_2\cos c_3,\\
g_2&=\sin c_1\sin c_2\sin c_3,\\
g_3&=2(\cos^2 c_1+\cos^2 c_2+\cos^2 c_3)-3.
\endaligned
\end{eqnarray}
To find the corresponding point $[c_1, c_2, c_3]$ of this quantum
circuit in the geometric representation, we only need to equate
Eqs.~\eqref{eq:4} and~\eqref{eq:24}, and thereby obtain:
\begin{eqnarray}
  \label{eq:6}
  \aligned
c_3&=0,\\
\cos c_1\cos c_2&
=\cos r_1 \cos r_2-\sin r_1 \sin r_2 \cos\beta_1 \cos\beta_2,\\
\cos^2c_1+\cos^2c_2&=(\cos\beta_1+\cos\beta_2)^2\sin^2\gamma_1\sin^2\gamma_2
+\cos^2\gamma_1+\cos^2\gamma_2\\
&\quad -2\cos\beta_1\cos\beta_2\sin\gamma_1\sin\gamma_2\cos(\gamma_1-\gamma_2).
\endaligned   
\end{eqnarray}
After some algebraic derivations, we obtain the following equations for
the tuning parameters $\beta_1$ and $\beta_2$:
\begin{eqnarray}
  \label{eq:7}
\aligned
  \cos\beta_1+\cos\beta_2&=\frac{\sqrt{\cos^2c_1+\cos^2c_2
-\cos^2\gamma_1-\cos^2\gamma_2+2(\cos\gamma_1\cos\gamma_2-\cos c_1\cos
c_2)\cos(\gamma_1-\gamma_2)}}{\sin\gamma_1\sin\gamma_2}\\
\cos\beta_1\cos\beta_2&=\frac{\cos\gamma_1\cos\gamma_2-\cos c_1
\cos c_2}{\sin\gamma_1\sin\gamma_2}
\endaligned
\end{eqnarray}
It is clear that $\cos\beta_1$ and $\cos\beta_2$ can be viewed as two roots
of the following quadratic equation:
\begin{eqnarray}
  \label{eq:8}
\aligned
  f(x)&=\sin\gamma_1\sin\gamma_2 x^2 +
(\cos^2c_1+\cos^2c_2-\cos^2\gamma_1-\cos^2\gamma_2\\
&\quad +2(\cos\gamma_1\cos\gamma_2-\cos c_1\cos c_2)\cos(\gamma_1-\gamma_2))^{\frac12}x
+\cos\gamma_1\cos\gamma_2-\cos c_1\cos c_2=0
\endaligned
\end{eqnarray}
Since $\gamma_1$, $\gamma_2\in (0, \frac\pi{2}]$, we have
$\sin\gamma_1\sin\gamma_2> 0$. To guarantee the existence of two roots
in the interval $[-1, 1]$, we need the following three conditions to
be satisfied: $f(1)\ge 0$, $f(-1) \ge 0$, and $\Delta\ge 0$, where
$\Delta$ is the discriminant of quadratic equation. It is not hard to
see that the first two conditions $f(1)\ge 0$ and $f(-1)\ge 0$ are
equivalent to the following inequality:
\begin{eqnarray}
  \label{eq:30}
\aligned
(\sin\gamma_1\sin\gamma_2
+\cos\gamma_1\cos\gamma_2-\cos c_1\cos c_2)^2
\ge &\cos^2c_1+\cos^2c_2-\cos^2\gamma_1-\cos^2\gamma_2\\
&+2(\cos\gamma_1\cos\gamma_2-\cos c_1\cos c_2)\cos(\gamma_1-\gamma_2).
\endaligned
\end{eqnarray}
After some algebraic derivations, Eq.~\eqref{eq:30} can be simplified
to $\sin^2c_1\sin^2c_2\ge 0$, which always holds true. Therefore, the
conditions $f(1)\ge 0$ and $f(-1)\ge 0$ are automatically satisfied
for any parameters $\beta_1$ and $\beta_2$. For the third condition,
we have
\begin{eqnarray}
  \label{eq:10}
  \Delta=(\cos c_1\cos(\gamma_1+\gamma_2)-\cos
  c_2)^2-\sin^2(\gamma_1+\gamma_2) \sin^2c_1.
\end{eqnarray}
To ensure $\Delta\ge 0$, we only need that $0\le c_1+c_2\le
\gamma_1+\gamma_2$. Therefore, all the local equivalence classes that
can be generated by these two Controlled-$U$ gates and local gates can
be described as $e^{c_1\frac{i}2\sigma_z^1\sigma_z^2}\cdot
e^{c_2\frac{i}2\sigma_z^1\sigma_z^2}$, where $0\le c_1+c_2\le
\gamma_1+\gamma_2$.
\end{proof}
\end{widetext}

\end{document}